\documentclass[british,english,english,twocolumn,showpacs]{revtex4}
\usepackage{graphicx}

\makeatletter

\makeatletter
\usepackage{color}
\usepackage{dcolumn}
\usepackage{babel}
\makeatother
\begin{document}

\title{The role of weak charging in metastable colloidal clusters}

\author{Christian L. Klix$^{\ast}$, Ken-ichiro Murata, and Hajime Tanaka$^{\dag}$}


\affiliation{Institute of Industrial Science, University of Tokyo, Meguro-ku,
Tokyo 153-8505, Japan.}


\author{Stephen R. Williams}

\affiliation{Research School of Chemistry, The Australian National University,
Canberra, ACT 0200, Australia.}

\author{Alex Malins and C. Patrick Royall$^{\ddag}$}


\affiliation{School of Chemistry, University of Bristol, Bristol BS8 1TS, UK.}

\date{May 20th, 2009}

\begin{abstract}
We study metastable clusters in a colloidal system with competing
interactions. A short-ranged polymer-induced attraction drives clustering,
while a weak, long-ranged electrostatic repulsion prevents extensive
aggregation. We compare experimental yields of cluster structures
expected from theory, which assumes simple addition of the competing
isotropic interactions. For clusters of size $4\leq m\leq6$, the
yield is significantly less than that expected. We attribute this
to an anisotropic self-organized surface charge distribution linked
to the cluster symmetry: non-additivity of electrostatic repulsion
and polymer-induced attraction. 7-membered clusters have a clear
optimal yield of the expected pentagonal bipyramid structure as a
function of strength of the attractive interaction. 
\end{abstract}


\maketitle

Clusters are intrinsically distinct from bulk matter,
with the restricted degrees of freedom making their energy landscapes
tractable, providing ideal model systems for studying kinetic pathways
\cite{wales}. Although atomic clusters have been extensively studied, 
direct visualization is challenging and often
limited to states of low temperature \cite{li2008}. Colloidal dispersions
provide a new medium by which to study clusters, and their well-defined
thermodynamic temperature allows analogy with atomic, molecular, and 
protein clusters. 
Furthermore, the ability to track particle coordinates in
3D yields detailed static and dynamic information usually only accessible
to computer simulation \cite{vanblaaderen1995}.

\begin{figure}
\begin{centering}
\includegraphics[width=8.5cm]{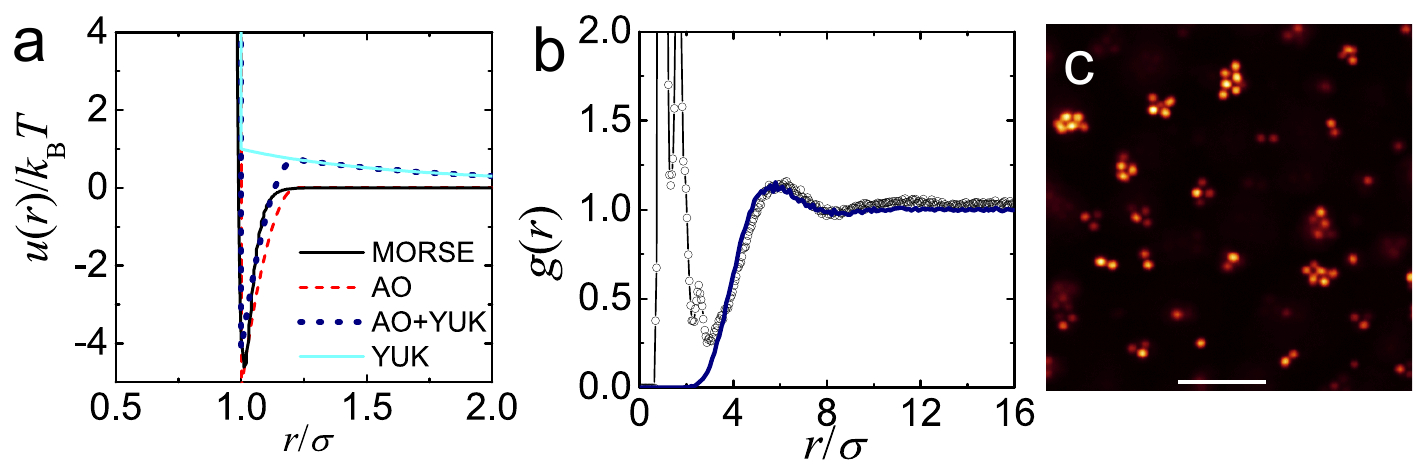} 
\par\end{centering}

\caption{(color online) (a) Various potentials: 
Morse (black line) denotes a Morse potential for a well depth 
$\varepsilon_{\rm M}=4.6$
and $\rho_{0}=33.1$.
AO (red dashed line) denotes Asakura-Oosawa
for $c_{\rm p}$$=6.02\times10^{-4}$, 
YUK (light cyan line)
denotes a Yukawa interaction
for a contact potential of $k_{\rm B}T$ and inverse Debye screening length
$\kappa\sigma=0.5$. 
AO+YUK (blue dotted line) sums AO and YUK illustrates a treatment of the
electrostatic repulsions as a small perturbation
(see text). (b) Determining the repulsive
interactions between the clusters. Experimental $g(r)$ for 
$c_{\rm p}=6.02\times10^{-4}$
(circles) and MC simulation (solid line) treating clusters
as single particles, each with a charge $Z=61$. (c) Confocal micrograph
with $c_{\rm p}=1.43\times10^{-3}$. Bar is $10$ $\mu$m. \label{figG}}
\end{figure}

Colloidal dispersions feature novel cluster phases \cite{segre2001,stradner2004,campbell2005,hong2006},
which open the possibility to synthesize colloidal clusters and `molecules'
\cite{manoharan2003,wilber2007,noya2007,anthony2008,zerrouki2008}
presenting exciting possibilities for new materials \cite{glotzer2007}.
To fully realize the potential of colloidal clusters and `molecules',
it is necessary to understand their behavior, and in particular to
investigate ways to optimize yields of desired structures \cite{wilber2007}.
Perhaps the simplest cluster-forming system is built around a spherically
symmetric attraction which may be realized by adding non-adsorbing
polymer to a colloidal suspension. The polymer osmotic pressure then
leads to an effective attraction between the colloids known as the
Asakura-Oosawa (AO) potential, whose strength is approximately proportional
to the polymer concentration $c_{\rm p}$ [Fig. \ref{figG}(a)].

While colloidal systems with purely attractive interactions may aggregate,
stabilization can be achieved with longer-ranged electrostatic repulsions
\cite{stradner2004}, which are often treated with a Yukawa (YUK)
potential \cite{royall2006}. Here we combine the competing short-ranged
attractions and weak long-ranged repulsions in colloidal systems with
the well-understood behavior of clusters interacting via short-ranged
attractions \cite{doye1995}. We neglect
possible many-body interactions \cite{lowen1998}.
As Fig. \ref{figG}(a) shows, for short 
interaction ranges the AO potential is similar to the Morse potential, 
whose range is set by a parameter $\rho_{0}$.
Now the structures of the global energy minima for clusters 
of Morse particles are known \cite{doye1995}, and for small clusters of
size $m<8$, the topology of this minimum structure is not sensitive to the range of the interaction,
nor to weak, long-ranged Yukawa repulsions \cite{mossa2004}. With
the exception of $m=6$ (see below), the global minimum of small clusters
in our colloidal system should therefore be identical to those tabulated
for the Morse interaction \cite{doye1995}. Using a relatively dilute
colloidal dispersion, we consider each cluster as a separate subsystem,
and investigate the cluster structure at the single particle level.
Using the topological cluster classification algorithm \cite{royall2008gel},
we compare the cluster structures found with those corresponding to 
Morse global energy minima, as a function
of polymer concentration (strength of the attractive interaction) \cite{doye1995,mossa2004}.
As we shall see, even in our simple model system, the behavior is
not as straightforward as one might expect. We believe this is due
to a breakdown of the Yukawa description of the electrostatic interactions.

We used poly(methyl methacrylate) (PMMA) colloids sterically stabilized
with polyhydroxyl steric acid. The colloids were labeled with fluorescent
rhodamine dye and had a diameter of $\sigma=2.0$ $\mu$m and a polydispersity
of $4$\% as determined with static light scattering. The van der
Waals interactions are neglected by closely matching the refractive
index between the solvent and particles, sedimentation is avoided
by matching the density. The colloid volume fraction was $\phi\approx0.02$.
We use a solvent composition of approximately $0.373$ cis decalin,
$0.093$ cyclohexyl bromide and $0.533$ tetrachloroethylene by mass,
in which the dielectric constant $\varepsilon_{\rm R}\approx2.2$. This
is a good solvent for the polymer, polystyrene (PS). We used a molecular
weight of $3.0\times10^{7}$, where our measurements of the interactions
agreed well with the Asakura-Oosawa (AO) model with some swelling
(35\%) of the polymer \cite{royall2007}, leading to a polymer-colloid
size ratio $q\approx0.22$ which corresponds to a Morse potential
with $\rho_{0}=33.1$ \cite{noro2000}. We studied the system
using confocal microscopy at the single particle level, using a Leica
SP5, with which we track the particle
coordinates to an accuracy of around $100$ nm. We impose a bond length of
$1.4\sigma$, this slightly larger value than the range of the attractive interaction
reflects particle tracking errors and polydispersity. Our results are 
robust to reasonable changes in the bond length. We sampled typically
$10^{5}$ coordinates per state point. No change in the cluster populations
was observed on the timescale of the experiments (1 day).

\begin{figure}
\begin{centering}
\includegraphics[width=8cm]{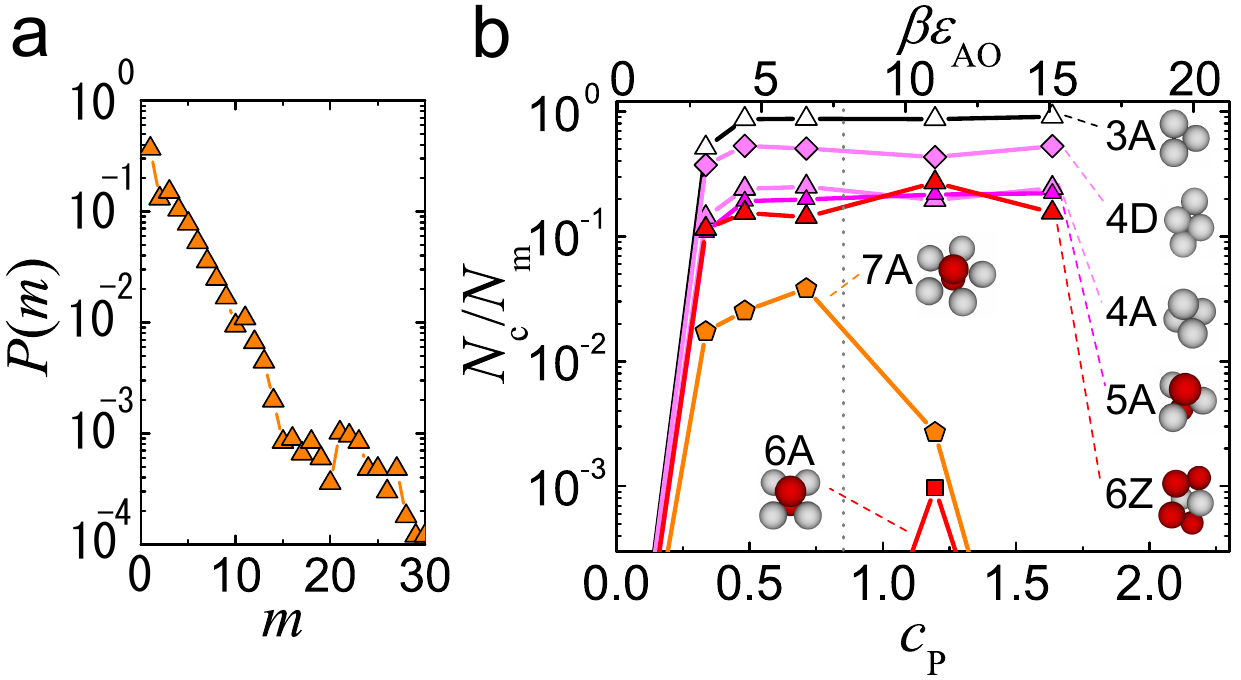} 
\par\end{centering}

\caption{\label{figIsolatedPopulations} (color online) (a) Cluster size distribution
for $c_{\rm p}$$=6.02\times10^{-4}$. (b) `Structural yield' $N_{c}/N_{m}$
as a function of $c_{\rm p}$. Only in the case of $m=3$ does the yield
of the global energy minimum (3A) approach $100$\%. Vertical dashed
line denotes polymer overlap. $\beta \varepsilon_{\rm AO}$ is the contact
potential for the AO interaction.}
\end{figure}

We find the order of magnitude of the colloid charge
in cluster fluid by treating the clusters as individual
particles and neglecting their different sizes \cite{sciortino2004}.
The size distribution is shown in Fig. \ref{figIsolatedPopulations}(a)
for a polymer weight fraction $c_{\rm p}=6.02\times10^{-4}$ the mean
$<m>=3.3$ and standard deviation is $4.4$. In Fig. \ref{figG}(b), we fit 
the radial distribution function $g(r)$ with Monte Carlo simulation 
according to a Yukawa potential \cite{royall2006}, yielding $Z=61\pm20$
charges per cluster, and a Debye screening length of 
$\kappa^{-1}=2.0\pm0.4\sigma$.
This suggests around $10-20$ charges per colloid, which corresponds
to a Yukawa potential at contact of 
$\beta\varepsilon_{\rm YUK}=Z^{2}l_{\rm B}/[(1+\kappa\sigma/2)^{2}\sigma]\sim1-3$
where 
$\beta=1/k_{\rm B}T$ ($k_{\rm B}$: Boltzmann's constant)
$l_{\rm B}\approx25$ nm is the Bjerrum length.
Although this charge is very low, similar values have been measured
for PMMA in apolar solvents \cite{roberts2007}. A confocal microscopy
image of the system is shown in Fig. \ref{figG}(c), we see 
the clusters are well isolated, consistent with our assumption that
each cluster may be treated as a separate system. Increasing 
$c_p$ promotes clustering.

We now proceed to consider the behavior of the individual clusters
(as a separate system). Figure \ref{figIsolatedPopulations}(b) shows
the `structural yield' $N_{c}/N_{m}$ as a function of the attractive
interaction $c_{\rm p}$. $N_{c}$ is the number of clusters of a given 
structure, $N_{m}$ is the total number of clusters containing $m$
colloids. For the smaller cluster sizes a greater portion are found
in the Morse global minimum, with the exception of the octahedal 6A
cluster which is much less popular than either the triangular bipyramid
5A or the pentagonal bipyramid 7A. Considering the topology of the
3A, 4A, 5A, and 6A clusters we see that the distance
between all the nearest neighbors will be the same due to the short
range nature of the AO potential and next nearest neighbors do not
contribute, if we ignore the weak electrostatic repulsion. The minimum
energy for these clusters is thus approximately proportional
to the number of nearest neighbors or contacts. However,
for $m=6$, global minimum for the Dzugutov potential \cite{doye2001}
has a $C_{2v}$ point group symmetry and the same number of contacts
as 6A. We therefore seek the this 6Z cluster using the TCC.

Neglecting the weak charge, both 6A and 6Z clusters have the same
ground state potential energy but there are other contributions to
the cluster's free energy: configurational
entropy and electrostatic energy. Indistinguishablility arguments
suggest that 6Z should be 12 times as popular as 6A. Moreover, 6Z
is further favored by its larger radius of gyration, which lowers
electrostatic energy \cite{groenewold2001}. As we shall see below,
the effects even of these weak electrostatic interactions can be rather
complex. However, the very large population difference in this system
is suggestive of a further mechanism at play: the kinetic pathway
and the associated energy barrier. Addition of particles leads to
the sequence 3A$\rightarrow$4A$\rightarrow$5A$\rightarrow$6Z while
the formation of 6A requires a bond to be broken and insertion of
a particle to form the 4-membered square ring.

\begin{figure}
\begin{centering}
\includegraphics[width=7.5cm]{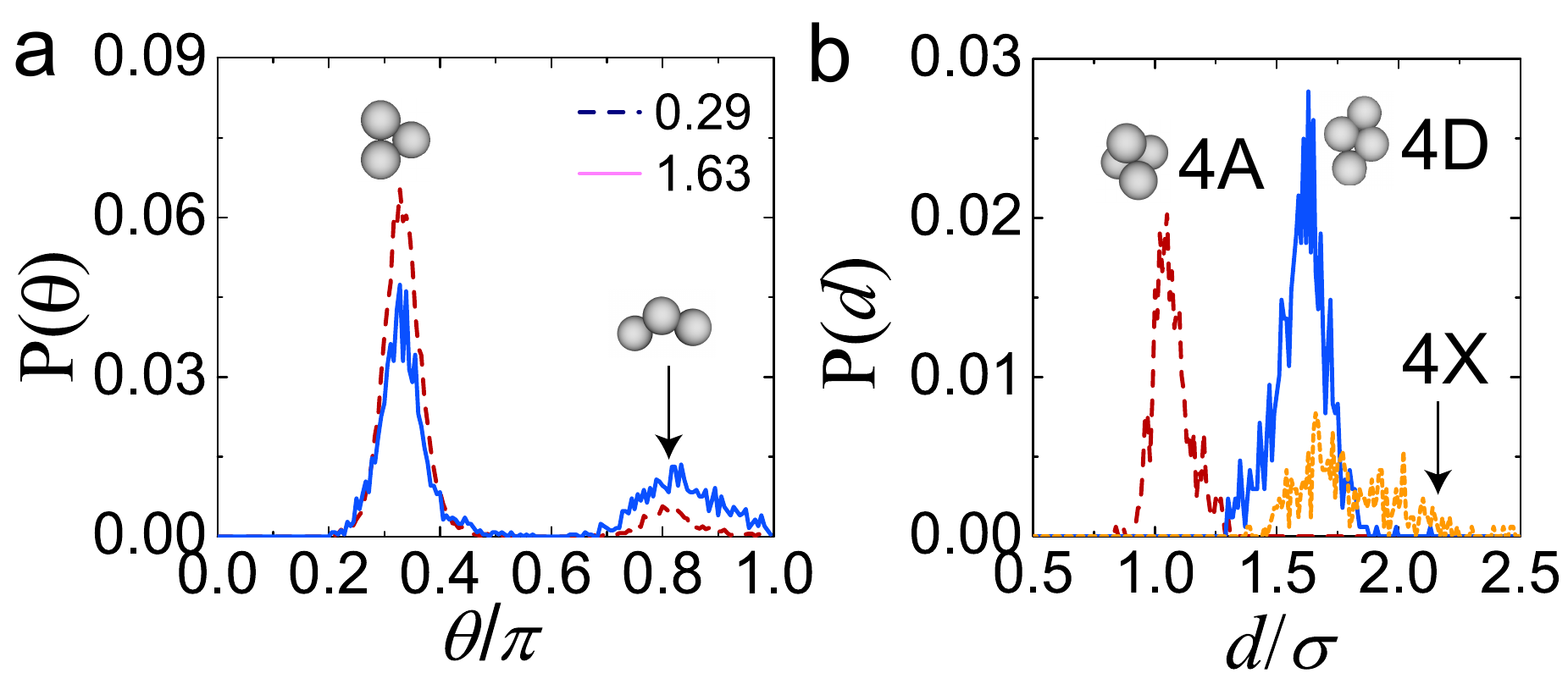} 
\par\end{centering}

\caption{\label{figAngles} (color online) (a) Distribution of bond angles
$\theta$ for $m=3$ clusters. As $c_{\rm p}$ (listed $\times10^{-3}$)
is increased, the population of 3A triangles increases. (b) Maximum
center separation, $m=4$. Red dashed line denotes 4A tetrahedron,
blue (solid) line 4D diamond and orange (dotted) line (`4X') higher
energy (fewer bonds). $c_{\rm p}=7.1\times10^{-4}$. }
\end{figure}

We expect an increase of polymer concentration (lowering of the effective
temperature) to promote clusters which maximize the number of bonds,
\emph{i.e.} 3-7A, or 6Z. This we see in Fig. \ref{figIsolatedPopulations}(b)
for $m=3$, in the near 100\% yield of 3A triangles at high $c_{\rm p}$.
In Fig. \ref{figAngles} (a) we show the distribution of bond angles
$\theta$ in $m=3$ at different polymer concentrations. In addition
to a higher 3A triangle population at higher $c_{\rm p}$, we see a peak
at large bond angle, showing that an elongated structure is preferred,
and that the system seems to sit either in a triangle or the linear
structure. We attribute this two-state behavior to the fact that the
colloids carry a (weak) charge, promoting elongated states \cite{groenewold2001}.
In the case of more complex clusters, such as 7A, an initial increase
in yield with $c_{\rm p}$ is followed by a decrease, presumably due to
frustration, at high interaction strength.

\begin{figure}
\begin{centering}
\includegraphics[width=7.5cm]{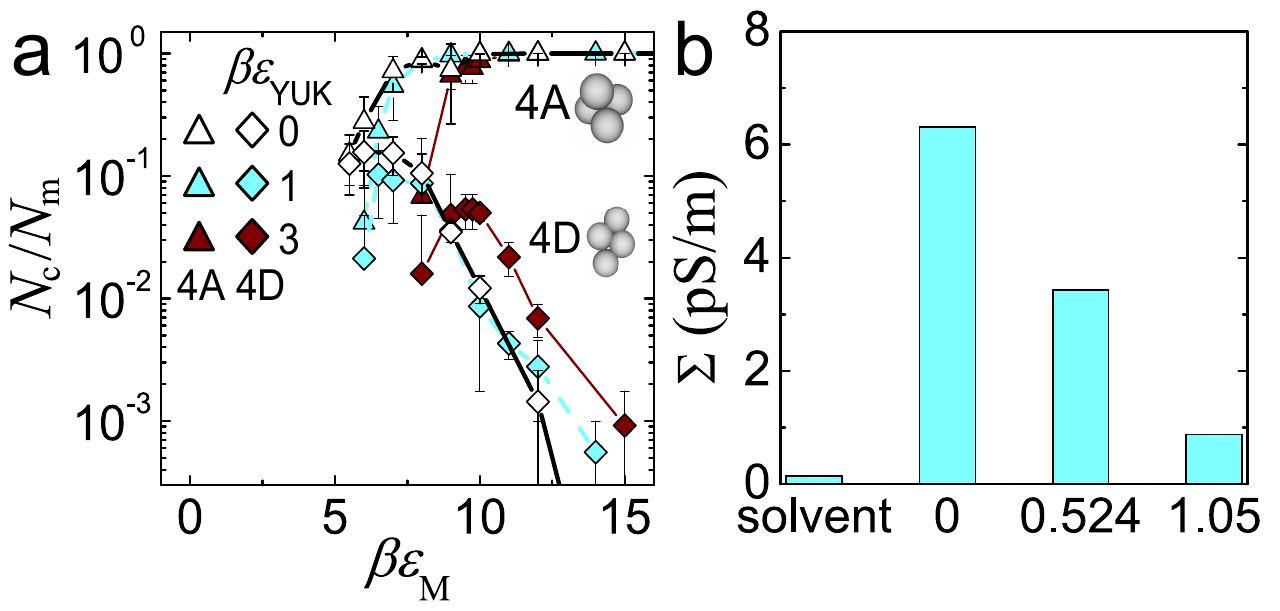} 
\par\end{centering}

\caption{\label{figIsolatedPopulationsBD} (color online) (a) Structural yields
$N_{c}/N_{m}$ as a function of well depth $\beta\varepsilon_{\rm M}$calculated
from Brownian dynamics simulations for the Morse potential for $\rho_{0}=33.1$
for $m=4$. Diamonds denote 4D and triangles 4A. Thick lines with
unfilled symbols denote Yukawa contact potential $\beta\varepsilon_{\rm YUK}=0$,
light blue (dashed) $\beta\varepsilon_{\rm YUK}=1$ and thin crimson lines
and filled symbols $\beta\varepsilon_{\rm YUK}=3$ and inverse Debye screening
length $\kappa\sigma=0.5$. (b) Conductivity of the solvent, and colloidal
suspension at various polymer concentrations $\times10^{-3}$.}
\end{figure}

However, intermediate cluster sizes, $4\leq m\leq6$ show a surprising
behavior, with little sensitivity to polymer concentration. Motivated
by our analysis of $m=3$ clusters, we consider structures in addition
to the ground state 4A tetrahedron in the $m=4$ population. Here
we consider the largest separation of colloid centers $d$. Now $d\approx\sigma$
for 4A tetrahedra, while the maximum value is $d\approx3\sigma$ for
a line of 4 particles. Figure \ref{figAngles}(b) shows two dominant
structures in the $m=4$ population: the 4A tetrahedra and a diamond-shaped
structure we term 4D, which is distinct from a square structure for
which $d=\sqrt{2}\sigma$. Unlike more complex (e.g. 7A) clusters,
for $m=4$, there is no geometric frustration in the formation of
tetrahedra. In fact Brownian dynamics simulations of four particles
interacting via the Morse potential \cite{malins2009} show a clear
peak in the 4D population, and a tendency to form tetrahedra at strong
interaction strength {[}Fig. \ref{figIsolatedPopulationsBD}(a)].
Furthermore, using an \textit{isotropic} Yukawa interaction with parameters
similar to those found in our experiments has little effect: 
we see a high yield of 4A at large $\beta\varepsilon_M$,
consistent with \cite{mossa2004}.

Why then do the experiments have such a high 4D population? One argument
might be that the AO model is starting to break down as we exceed
overlap, which occurs at $c_{\rm p}\approx8.5\times10^{-4}$ as indicated
in Fig. \ref{figIsolatedPopulations}. However for this size ratio
we expect only a slight effect in the AO interactions at overlap 
\cite{louis2002}.
Instead, we postulate that 4D is stabilized by the weak charging present
in this system, which unlike the spherically symmetric distribution
often assumed, may be anisotropically distributed on colloid surfaces,
leading to anisotropic electrostatic interactions \cite{hong2006}.

While electrostatic charging mechanisms in these non-aqueous systems
are not yet fully understood, it is reasonable to suppose that charging
may be related to the free colloid surface, which may be reduced upon
clustering. To this end, we measured the conductivity of the system
as a function of $c_p$ with dielectric spectroscopy
[Fig. \ref{figIsolatedPopulationsBD}(b)]. As we do not add salt,
we assume the system is in the counter-ion dominated regime, and we
note a large increase in conductivity relative to the pure solvent.
What is evident from Fig. \ref{figIsolatedPopulationsBD}(b) is that
the introduction of the colloids massively increases the conductivity
relative to the pure solvent, and that adding polymer then decreases
the conductivity. We associate this reduction in ionic strength (charge
per colloid) with clustering, and argue that the charge is indeed
associated with the free surface of the colloids, which is reduced
upon clustering. While precise determination of the ionic strength
from conductivity measurements requires detailed knowledge of the
ionic species present, assuming Walden's rule and a typical limiting
molar conductance for ions in these systems of $4$ cm$^{2}$mol$^{-1}$\cite{royall2006}
we arrive at an ionic strength around $10^{-10}$ M, consistent with
the Debye length $\kappa^{-1}\sim2\sigma$ found with our MC simulations
[Fig. \ref{figG}(b)].

4D and 4A clusters are formed by an attachment of one particle to
3A. For 3A$\rightarrow$4A there are three direct contact points to
be formed, whereas for 3A$\rightarrow$4D there are only two. Thus
4D may be more easily formed than 4A due to a reduced electrostatic
barrier and smaller loss of translational entropy of counter ions.
We note that the latter factor is not considered at all in a model
with a Yukawa potential. Furthermore, we stress that a Yukawa potential
should break down at short distances and thus fail in describing aggregation
of particles. For 4D $d\sim\sqrt{3}\sigma$ whereas for 4A $d\sim\sigma$.
So 4D is more favored than 4A electrostatically. We note that although
there is no structural frustration in the 4D$\rightarrow$4A transition,
there is only one path, if no bonds are broken; one or both of the two end
particles must `roll' around the central two. We note that the 4D-4A transformation,
always leads to a decrease in $d$, which may be prohibited by electrostatic repulsions.
Thus, once 4D is formed, the transformation from 4D to 4A is suppressed: 
kinetic arrest. That 4D formation is promoted
and 4D$\rightarrow$4A transformation is suppressed due 
may lead to the high fraction of 4D. It is possible that the
charging mechanism leads to a scenario in which the equilibrium
populations are a distribution dominated by both 4D and 4A as found.
However, the lack of response of the system to increasing the strength
of attraction by a factor of three leads us to the conclusion that
some form of kinetic arrest is more likely. On the other hand, our
results suggest that it is not so difficult to to make a 3A triangle
from a linear cluster. Although the reason for the difference is not
immediately clear, we note that the linear$\rightarrow$3A transition
is not geometrically constrained to follow a single path.

Here we consider some microscopic aspects of this problem. The low
surface charge density (small $Z$) implies that the discrete nature of 
charges may be important. Furthermore, the recombination rates of ions 
might be very slow: even for small ions, diffusion-controlled 
reaction rate constants
are $10^{9}$-$10^{10}$ dm$^{3}$M$^{-1}$s$^{-1}$, of order seconds
or tens of seconds here. For the larger ions or complexes relevant
to these apolar solvents \cite{royall2006,roberts2007} the rate may
be lower still. That the charges are discrete in both space
and time may play an important role in the above-described frustration.
Further work is necessary for elucidating the exact mechanism responsible
for the observed cluster type distribution. 

Our study suggests that the one-component description is breaking
down, even for these simple colloidal systems. In the case of $m=6$,
two structures with similar energies compete, however, the 6Z is very
much more popular than the 6A octahedron. Although equilibrium arguments
favor this trend we note that 6Z is structurally similar to 5A, and
thus may likely be kinetically favored. The yield of the fivefold
symmetric 7A shows signs of frustration at relatively deep quenches,
reminiscent of patchy particles \cite{wilber2007}. Our results form
the beginnings of an understanding of self-assembly of colloidal molecules 
as a function of attraction strength; we hope they will stimulate the 
development of more powerful models. 

The authors are grateful to Wuge Briscoe, Didi Derks, Jeroen van Duijneveldt, Rob Jack and Matthias Schmidt
for a critical reading of the manuscript. CLK thanks the Deutscher
Akademischer Austauschdienst for financial assistance. CPR acknowledges
the Royal Society for financial support. HT acknowledges a Grant-in-Aid
from the Ministry of Education, Culture, Sports, Science and Technology,
Japan.

\expandafter\ifx\csname natexlab\endcsname\relax\def\natexlab#1{#1}\fi
\expandafter\ifx\csname bibnamefont\endcsname\relax \def\bibnamefont#1{#1}\fi
\expandafter\ifx\csname bibfnamefont\endcsname\relax \def\bibfnamefont#1{#1}\fi
\expandafter\ifx\csname citenamefont\endcsname\relax \def\citenamefont#1{#1}\fi
\expandafter\ifx\csname url\endcsname\relax \def\url#1{\texttt{#1}}\fi
\expandafter\ifx\csname urlprefix\endcsname\relax\def\urlprefix{URL }\fi
\providecommand{\bibinfo}[2]{#2} \providecommand{\eprint}[2][]{\url{#2}}

\end{document}